\documentclass[aps,pre,10pt,twocolumn,byrevtex,superscriptaddress,footinbib,bibnotes,showpacs,showkeys,longbibliography]{revtex4-2}
\usepackage{microtype}
\usepackage{amssymb}
\usepackage{amsmath}
\usepackage{graphicx}
\usepackage{color}
\usepackage{psfrag}

\definecolor{myblue}{rgb}{0.153,0.322,0.706}
\usepackage[colorlinks,linkcolor=myblue,urlcolor=myblue,citecolor=myblue,breaklinks=true]{hyperref}

\newcommand{\be}{\begin{equation}}
\newcommand{\ee}{\end{equation}}
\newcommand{\ra}{\rightarrow}
\newcommand{\reals}{\mathbb{R}}

\newcommand{\tA}{\tilde A}
\newcommand{\tX}{\tilde X}
\newcommand{\rev}{\textrm{rev}}
\newcommand{\irr}{\textrm{irr}}
\newcommand\at[2]{\left.#1\right|_{#2}}
\newcommand{\p}{\partial}
\newcommand{\hj}{\hat\jmath}

\newcommand{\Itwo}{I^{(2)}}
\newcommand{\Itwofive}{I^{(2.5)}}
\newcommand{\sprod}[2]{\left\langle #1, #2\right\rangle_{\rho D}}
\newcommand{\sprodd}[2]{\left\langle #1, #2\right\rangle}
\newcommand{\snorm}[1]{\left\| #1 \right\|_{\rho D}}
\newcommand{\transp}{\mathsf{T}}
\begin{document}
\title{Role of current fluctuations in nonreversible samplers}

\author{Francesco Coghi}
\email{francesco.coghi@gmail.com}
\affiliation{School of Mathematical Sciences, Queen Mary University of London, London E1 4NS, England}

\author{Rapha\"el Chetrite}
\email{Raphael.Chetrite@unice.fr}
\affiliation{\mbox{Laboratoire J.~A. Dieudonn\'e, UMR CNRS 7351, Universit\'e de Nice Sophia Antipolis, Nice 06108, France}}

\author{Hugo Touchette}
\email{htouchette@sun.ac.za, htouchet@alum.mit.edu}
\affiliation{Department of Mathematical Sciences, Stellenbosch University, Stellenbosch 7600, South Africa}
\date{\today}

\begin{abstract}
It is known that the distribution of nonreversible Markov processes breaking the detailed balance condition converges faster to the stationary distribution compared to reversible processes having the same stationary distribution. This is used in practice to accelerate Markov chain Monte Carlo algorithms that sample the Gibbs distribution by adding nonreversible transitions or non-gradient drift terms. The breaking of detailed balance also accelerates the convergence of empirical estimators to their ergodic expectation in the long-time limit. Here, we give a physical interpretation of this second form of acceleration in terms of currents associated with the fluctuations of empirical estimators using the level 2.5 of large deviations, which characterises the likelihood of density and current fluctuations in Markov processes. Focusing on diffusion processes, we show that there is accelerated convergence because estimator fluctuations arise in general with current fluctuations, leading to an added large deviation cost compared to the reversible case, which shows no current. We study the current fluctuation most likely to arise in conjunction with a given estimator fluctuation and provide bounds on the acceleration, based on approximations of this current. We illustrate these results for the Ornstein--Uhlenbeck process in two dimensions and the Brownian motion on the circle.
\end{abstract}

%\keywords{Markov processes, Langevin dynamics, nonreversible processes, large deviations}

\maketitle

%%%%%%%%%%%%%%%%%%%%%%%%%%%%%%%%%%%%%%%%%%%%%%%%%%%%%%%%%%%
\section{Introduction}

Markov chain Monte Carlo algorithms, such as the Metropolis--Hastings algorithm and the Langevin sampler, are commonly used in physics, chemistry, and statistics to sample a distribution $\pi$ by simulating an ergodic Markov process $X(t)$ whose stationary distribution is $\pi$ \cite{newman1999,liu2001,asmussen2007,kalos2008}. In recent years, it has been shown in many works \cite{hwang1993,hwang2005,lelievre2013,wu2014,turitsyn2011,ichiki2013,ottobre2016,sakai2016,suwa2010,fernandes2011,chen2013c,hwang2015,duncan2016,duncan2017} that the convergence of such algorithms can be improved by modifying them so as to break the detailed balance condition, thus rendering them nonreversible, while preserving $\pi$ as the stationary distribution. For algorithms based on Markov chains, this is achieved by adding transitions between states in order to create cycles, while for algorithms based on diffusion equations, such as the Langevin sampler, this is achieved by adding non-gradient terms to  the drift. In either case, it is known that the time-dependent distribution $\pi_t$ of $X(t)$ converges faster to $\pi$, as the gap in the spectrum of the generator of $X(t)$ is increased compared to its reversible version \cite{hwang1993,hwang2005,lelievre2013,wu2014,turitsyn2011,ichiki2013,ottobre2016,sakai2016}, a result already noted in the 1970s by Risken \cite{risken1972}. 

The improved convergence of $\pi_t$, or any quantities derived from this distribution, means computationally that the mixing or ``burn in'' time needed for a Markov process to start sampling according to $\pi$, starting from some initial distribution $\pi_0$, is reduced by forcing it to be nonreversible. In a more fundamental way, it is also known that statistical estimators based on nonreversible processes have better convergence properties, as their variance is reduced by breaking detailed balance \cite{chen2013c,hwang2015,duncan2016,duncan2017}, which means that the simulation time needed for an estimator to reach its ergodic value within some fixed threshold or error bar is also reduced. This is important in practice as most quantities estimated from Monte Carlo simulations (e.g., susceptibilities, transport coefficients, model parameters, etc.) take the form of time averages \cite{lelievre2016} having a bias or systematic error, often related to the mixing time, and a statistical error, which is generally more important, determined by the estimator variance \cite{newman1999,liu2001,asmussen2007,kalos2008}.

In a series of papers \cite{bellet2015,bellet2015b,bellet2016}, Rey-Bellet and Spiliopoulos have shown that the improved convergence of estimators obtained with nonreversible samplers can be understood in an elegant way using the theory of large deviations \cite{dembo1998,hollander2000,touchette2009}. Assuming that the distribution of an estimator $A_T$ has a large deviation form with the integration time $T$, which is generally the case in Monte Carlo simulations \cite{bucklew1990}, they show that the rate function of $A_T$ obtained with a nonreversible process is always larger than the rate function obtained with a reversible process with the same ergodic distribution $\pi$, except at the ergodic value of $A_T$, corresponding to the minimum and zero of the rate function, which is the same for both processes. Since rate functions determine the likelihood of the fluctuations of $A_T$, this means that estimator fluctuations are exponentially suppressed in a nonreversible process, compared to reversible ones, leading to a faster convergence of $A_T$ to its ergodic value as $T\ra\infty$. Moreover, since the asymptotic variance of an estimator is given by the reciprocal of the second derivative of its rate function at its minimum \cite{touchette2009}, they obtain as a corollary that an increase in the rate function leads to a reduction of the estimator variance.

In this paper, we present a different derivation of these results for the case of diffusion equations, based on the large deviations of currents, which provides a clear physical explanation of why nonreversible processes are more efficient than reversible ones for sampling estimators. The new derivation builds on recent works on the level 2.5 of large deviations, which describes the likelihood of joint fluctuations of the empirical density and empirical current in Markov processes \cite{maes2008b,chernyak2014,barato2015,bertini2015,hoppenau2016}, in addition to the theory of effective processes \cite{jack2010b,chetrite2013,chetrite2014,jack2015,chetrite2015}, which provides a description of how fluctuations of time-averaged quantities arise in the long-time limit in terms of effective processes having modified rates or drift terms. 

Using these two formalisms, we show that estimator fluctuations are suppressed in nonreversible diffusions as a result of current fluctuations that incur an additional large deviation cost compared to reversible diffusions, which have no currents because of the detailed balance condition. In other words, estimator fluctuations in a nonreversible diffusion are created by an effective process in which the density as well as the current are modified, which is more unlikely than creating the same fluctuations via an effective process which is reversible and, therefore, only changes the density \cite{chetrite2014}.

This interplay between density and current fluctuations applies beyond Monte Carlo algorithms to any Markov processes and time-integrated functionals or ``observables'' of these processes, and can be used to determine how dynamical fluctuations of such observables, which correspond in physics to measurable quantities \cite{sekimoto2010,seifert2012,touchette2017}, arise from optimal density and current fluctuations \cite{chetrite2014,jack2015,chetrite2015}. Here, we determine the optimal density and current underlying an estimator fluctuation, and show how approximations of the optimal current can be used to obtain upper bounds on the rate function associated with the nonreversible process, similar mathematically to entropic bounds recently derived in the context of stochastic thermodynamics \cite{gingrich2016,pietzonka2015,gingrich2017,nardini2018,li2019}. 

We illustrate these results with two simple but classical stochastic processes, namely, the Ornstein--Uhlenbeck process in two dimensions with normal or transverse drift and the simple diffusion on the circle. Applications involving other Markov processes and observables are discussed in the conclusion.

%%%%%%%%%%%%%%%%%%%%%%%%%%%%%%%%%%%%%%%%%%%%%%%%%%%%%%%%%%%
\section{Model and problem}

The problem that we consider is to sample the following Gibbs distribution on $\reals^d$:
\be
\pi(x)=\frac{e^{-\beta U(x)}}{\int_{\reals^d} e^{-\beta U(x)}dx},
\label{eqgibbs1}
\ee
where $U:\reals^d\ra\reals$ is a potential function, such that $\pi$ is normalizable, and $\beta\in\reals$ is an inverse temperature parameter controlling the variance of $\pi$. For this purpose, we use two diffusion processes that have $\pi$ as their stationary distribution, so they can be simulated in time to obtain (correlated) samples from this distribution, which may then be used to estimate expectations of the form
\be
E_\pi[f(X)] = \int_{\reals^d} f(x) \pi(x)\, dx,
\label{eqexp1}
\ee
where $f:\reals^d\ra\reals$ is any (test) function with finite expectation with respect to $\pi$.

The first process that we consider is a gradient diffusion $X(t)\in \reals^d$ given by the following (It\^o) stochastic differential equation (SDE):
\be
dX(t) = -\nabla U(X(t)) dt+\sqrt{\frac{2}{\beta}}\, dW(t),
\label{eqsde1}
\ee
where $W(t)\in\reals^d$ is a vector of independent Brownian motions. It is known that this SDE defines a time-reversible process satisfying the detailed balance condition \cite{risken1996}, whose unique stationary (ergodic) density is $\pi$ in \eqref{eqgibbs1}.

The second process that we consider is a nonreversible perturbation of the gradient diffusion, defined by 
\be
dX(t) = [-\nabla U(X(t)) +C(X(t))]dt+ \sqrt{\frac{2}{\beta}}\, dW(t),
\label{eqsde2}
\ee
where $C:\reals^d\ra\reals^d$ is a smooth vector field satisfying the condition
\be
\nabla\cdot(C\pi)  = 0,
\label{eqerg1}
\ee 
which ensures that $\pi$ in \eqref{eqgibbs1} is a solution of the time-independent Fokker--Planck equation \cite{risken1996}, and so remains the stationary distribution. This diffusion is nonreversible in the sense that it violates for $C\neq 0$ the detailed balance condition with respect to $\pi$, leading to a non-zero stationary current field given by
\be
J_{F,\pi} = F\pi -\frac{D}{2}\nabla\pi = C\pi,
\label{eqcurr1}
\ee
having identified
\be
F=-\nabla U+C
\ee 
as the total drift of the nonreversible diffusion and
\be
D=\frac{2}{\beta}.
\ee 
as the noise variance. Many choices for $C$ preserving $\pi$ are possible: one can use, for example, $C=S\nabla U$, where $S$ is any antisymmetric matrix, or more generally a $C$ such that $\nabla\cdot C=0$ with $C\cdot \nabla U=0$. 

The advantage of using the nonreversible SDE, as mentioned in the introduction, is that it performs better as a sampler of $\pi$ than the reversible SDE. The precise notion of performance that we consider is the statistical performance mentioned before, related to the simulation time $T$ that one needs to use in a simulation in order for the time average
\be
A_T = \frac{1}{T}\int_0^T f(X(t)) dt
\label{eqobs1}
\ee
to converge to the expectation $E_\pi[f(X)]$ shown in \eqref{eqexp1} with a given confidence interval. The convergence of $A_T$, which is the natural estimator of  $E_\pi[f(X)]$, is guaranteed by the ergodic theorem, which states that
\be
A_T\ra a^*=E_\pi[f(X)]
\label{eqlln1}
\ee
in probability as $T\ra\infty$ \cite{asmussen2007}. From the central limit theorem, generalized to Markov processes, it is known that the fluctuations of $A_T$ around the limit value $a^*$ are approximately Gaussian with a variance that decreases with $T$ according to $\sigma^2/T$, where
\be
\sigma^2 = \lim_{T\ra\infty} T E[(A_T - a^*)^2] = \lim_{T\ra\infty} T\, \textrm{var}(A_T)
\ee
is the \emph{asymptotic variance} of $A_T$ \cite{jones2004,komorowski2012,cattiaux2012}. Hence, the smaller the asymptotic variance, the better it is statistically, as a smaller Gaussian confidence interval (viz., error bar) can be reached for a given simulation time $T$.

In this context, it has been shown in many studies \cite{chen2013c,hwang2015,duncan2016,duncan2017} that introducing a nonreversible drift preserving $\pi$ systematically improves the asymptotic variance. To be more precise, let $\sigma^2_{0}$ and $\sigma^2_{C}$ denote the asymptotic variances of the estimator $A_T$ obtained, respectively, with the reversible ($C=0$) and nonreversible dynamics. Then
\be
\sigma^2_{C}\leq \sigma^2_{0}
\label{eqvarbound1}
\ee
with equality, under some conditions, if and only if $C=0$. As a result, for a large but finite integration time $T$, the estimator $A_T$ calculated along a trajectory of the nonreversible dynamics will have a smaller error bar than if it is calculated along a trajectory of the reversible dynamics. This also holds when adding nonreversible transitions in Markov chains and Markov jump processes \cite{chen2013c}, and so implies overall that the statistical estimation of expectations is accelerated, in the asymptotic variance sense, by nonreversible Markov dynamics \footnote{It should be clear that this statistical acceleration of $A_T$ is different conceptually from the acceleration of the mixing time, related to the spectral gap of the generator of $X(t)$. To be sure, take $X(0)\sim \pi$. Then the mixing time is obviously 0 in both the reversible and nonreversible dynamics, yet $A_T$ still has different asymptotic variances with respect to these two dynamics, which means that it converges to $a^*$ with different speeds as $T\ra\infty$.}.

The bound \eqref{eqvarbound1} for the asymptotic variance can be derived using a probabilistic version of the Poisson equation \cite{chen2013c,hwang2015,duncan2016} or using large deviation techniques, as shown recently by Rey-Bellet and Spiliopoulos \cite{bellet2015,bellet2015b,bellet2016} (see also Bierkens \cite{bierkens2016}). The latter approach, which is the focus of this paper, is based on the fact that, for many processes and estimators $A_T$ of interest, the probability density $P_T(a)$ of $A_T$ decays as $T\ra\infty$ according to
\be
P_T(a) \approx e^{-TI(a)},
\ee
up to corrections that are sublinear in $T$ in the exponent, so the decaying exponential is the dominant term of $P_T(a)$. This approximation is called in large deviation theory the \emph{large deviation principle} (LDP) \cite{dembo1998,hollander2000,touchette2009}. The exponent $I(a)$ controlling that decay is called the \emph{rate function} and can be obtained by the limit
\be
\lim_{T\ra\infty} - \frac{1}{T}\ln P_T(a) = I(a),
\ee
which is a simplified version of the limit used in large deviation theory to define the LDP \cite{dembo1998}.

The rate function provides detailed information about the likelihood of the different values (viz., fluctuations) of $A_T$, and can be obtained without calculating $P_T(a)$ exactly, which explains why it is often used in simulations \cite{bucklew1990} and statistical physics \cite{touchette2009} to study the properties of stochastic processes. In particular, the ergodic value of $A_T$ is determined by noting that $I(a)$ is always positive and has, for ergodic processes,  a single zero located at $a^*$ \cite{dembo1998}. Thus, values of $A_T$ different from $a^*$ are exponentially unlikely with $T$, implying the limit of the ergodic theorem in \eqref{eqlln1}. In general, $I(a)$ also has a parabolic shape around its minimum $a^*$, related to the Gaussian nature of the small fluctuations of $A_T$ around $a^*$, which implies that the asymptotic variance can be obtained as \cite{bryc1993}
\be
\sigma^2=\frac{1}{I''(a^*)}.
\label{eqasympvarld1}
\ee

Applying large deviation theory to the sampling problem, Rey-Bellet and Spiliopoulos \cite{bellet2015,bellet2015b,bellet2016} proved that
\be
I_{C}(a) \geq I_{0}(a),
\label{eqldbound1}
\ee
where, similarly to \eqref{eqvarbound1}, $I_{0}$ and $I_{C}$ denote the rate functions of $A_T$ obtained with respect to the reversible and nonreversible processes defined before. Moreover, they showed that the two rate functions are equal, under some conditions, only for the value $a^*$, which is obviously such that $I_{C}(a^*) = I_{0}(a^*)=0$, since both processes have the same $\pi$. Using \eqref{eqasympvarld1}, they then recover the acceleration bound \eqref{eqvarbound1} for the asymptotic variance.

Our goal in the next sections is to derive the large deviation bound \eqref{eqldbound1}, which is obviously stronger than the variance bound \eqref{eqvarbound1}, using the so-called level 2.5 of large deviations, related to density and current fluctuations in Markov processes \cite{maes2008b,chernyak2014,barato2015,bertini2015}. For this purpose, we review in the next section the derivation of \eqref{eqldbound1} by Rey-Bellet and Spiliopoulos, who used the level 2 of large deviations describing density fluctuations, and then present in Sec.~\ref{seclevel25} the basis of the level 2.5 together with our proof of \eqref{eqldbound1} using the latter level. 

%%%%%%%%%%%%%%%%%%%%%%%%%%%%%%%%%%%%%%%%%%%%%%%%%%%%%%%%%%%
\section{Level-2 large deviations}
\label{seclevel2}

The proof of the large deviation bound \eqref{eqldbound1} given by Rey-Bellet and Spiliopoulos \cite{bellet2015} relies on the fact that the observable $A_T$, as defined in \eqref{eqobs1}, is purely additive in time and can therefore be expressed as
\be
A_T = \tA(\rho_T) = \int_{\reals^d} f(x)  \rho_T(x)\, dx,
\ee
where
\be
\rho_T(x) = \frac{1}{T}\int_0^T \delta (X(t)-x)dt.
\ee
The latter estimator represents the fraction of time that a trajectory spends in the state $x$ during the time interval $[0,T]$, and is called for this reason the empirical occupation or local time at $x$. It can be seen as a random function, which converges in the ergodic limit to the stationary distribution $\pi$. The fluctuations of $\rho_T$ around that concentration point are known to be described by an LDP, whose rate function was found by Donsker and Varadhan \cite{donsker1975} to be given, for ergodic processes, by 
\be
\Itwo_C(\rho) =-\inf_{u>0} \int_{\reals^d} \rho(x) \frac{Lu(x)}{u(x)}\, dx,
\label{eqdv1}
\ee
where
\be
L = F\cdot \nabla+\frac{D}{2}\nabla^2
\label{eqgen1}
\ee
is the generator of the nonreversible process and the minimization in \eqref{eqdv1} is over all functions $u$ that are positive. Given that $\rho_T$ has an LDP and that $A_T$ is a function of  $\rho_T$, then $A_T$ must also have an LDP with a rate function given by
\be
I_{C}(a) = \inf_{\rho:\tA(\rho)=a} I^{(2)}_C(\rho),
\label{eqcont1}
\ee
where the minimization is over all densities $\rho$ that are normalized and such that $\tA(\rho)=E_\rho[f(X)]=a$. 

This minimization, which is known in large deviation theory as the contraction principle \cite{dembo1998,hollander2000,touchette2009}, is the main result needed for proving \eqref{eqldbound1}. Its interpretation should be clear: among the many empirical distributions $\rho$ that can be observed as leading to or underlying a given fluctuation $A_T=a$, the most likely minimizes $\Itwo_C(\rho)$ given the constraint $\tA(\rho)=a$, so the probability of $A_T$ is given to dominant exponential order in $T$ by the probability of that constrained $\rho$. In large deviation theory, we say that the large deviations of $\rho_T$ are contracted down to those of $A_T$, which explains why $\Itwo_C$ is referred to as the level-2 rate function, the level 1 being the lower level of the large deviations of $A_T$ described by $I_C$ \cite{touchette2009}.

The minimization in \eqref{eqdv1} cannot be reduced, in general, to an explicit expression for $\Itwo_C(\rho)$; however, it can be manipulated, following \cite{bellet2015}, to show that
\be
\Itwo_C(\rho) = \Itwo_0(\rho)+\frac{D}{2}\int_{\reals^d} |\nabla \psi(x)-\nabla U(x)|^2 \rho(x)\, dx,
\label{eqrbs1}
\ee
where $\Itwo_0$ is the level-2 rate function obtained with $C=0$, $U$ is the potential associated with the invariant distribution $\pi$, and $\psi$ is the unique solution (up to a constant) of the following equation:
\be
\nabla\cdot [\rho(-\nabla U+C+\nabla\psi)]=0.
\label{eqrbscons1}
\ee
Since the second term in the right-hand side of \eqref{eqrbs1} is positive, we then have
\be
\Itwo_C(\rho)\geq \Itwo_0(\rho).
\label{eqldbound2}
\ee
As this bound holds for all normalised distributions $\rho$, we can use it in the contraction \eqref{eqcont1}, thus recovering the bound \eqref{eqldbound1} for the level-1 large deviations of $A_T$.

We refer to \cite{bellet2015} for a more complete presentation of this reasoning and for a discussion of the conditions implying equality in all the bounds. Further conditions must be imposed to derive the asymptotic variance bound in \eqref{eqvarbound1}, which are discussed in the same reference. 

For the remainder, it is important to note that $\Itwo_0$ has an explicit expression given by
\be
\Itwo_0(\rho) = \frac{D}{2}\int_{\reals^d} \left|\nabla \sqrt{\textstyle\frac{\rho(x)}{\pi(x)}}\right|^2 \pi(x)\, dx.
\label{eqldp2}
\ee
Thus, the level-2 rate function is explicit when dealing with gradient (reversible) diffusions \footnote{Reversible Markov chains evolving in continuous time also have an explicit level-2 rate function, but not reversible Markov chains evolving in discrete time \cite{hollander2000}.}, as found by Donsker and Varadhan \cite{donsker1975}, as well as by G\"artner \cite{gartner1977}. A different expression for the same rate function is
\be
\Itwo_0(\rho)=\frac{1}{2D}\int_{\reals^d} |\nabla U(x)-\nabla V(x)|^2 \rho(x)\, dx,
\label{eqldp22}
\ee 
where
\be
V(x) = -\frac{D}{2}\ln \rho(x)
\label{eqVpot1}
\ee 
is the potential associated with the distribution $\rho$ ($U$ is the potential associated with $\pi$). A physical interpretation of this formula will be given in the next section.

%%%%%%%%%%%%%%%%%%%%%%%%%%%%%%%%%%%%%%%%%%%%%%%%%%%%%%%%%%%
\section{Level-2.5 large deviations}
\label{seclevel25}

The fact that the rate function of the empirical distribution $\rho_T$ is not explicit in general arises essentially because an ergodic Markov diffusion is not uniquely determined by its stationary distribution alone -- many diffusions having different stationary currents, and thus different nonreversible properties, can have the same stationary distribution, as is clear from the problem considered here. To uniquely identify the drift $F$ of a diffusion, for a given $D$, one must fix the stationary distribution $\pi$ \textit{and} the stationary current $J_{F,\pi}$, given in \eqref{eqcurr1}. This suggests that the large deviations of a fluctuating version of the current together with the empirical distribution $\rho_T$, which is the fluctuating version of $\pi$, might be described by a rate function which is explicit \footnote{A physical analogy can be drawn here: to describe an electrical system that has no current, we only need the charge density (electrostatics). If currents are present, then the charge density is insufficient --  we also need the current to describe the full system (electrodynamics).}. 

This is indeed the case, as was found recently in many studies coming mainly from statistical physics (see \cite{barato2015} for references) and is the main result underlying our proof of the level-1 bound \eqref{eqldbound1} and the associated bound \eqref{eqvarbound1} for the asymptotic variance. For completeness, we briefly review next this large deviation result, commonly referred to as the level 2.5 of large deviations, since it sits above or expands the level 2, for reasons that will become obvious, but sits below the level 3, referred to in large deviation theory as the process level \cite{touchette2009}. 

The fluctuating current that enters in the level 2.5 is defined formally as
\be
J_T(x) = \frac{1}{T}\int_0^T \delta (X(t)-x)\circ dX(t),
\ee
where $\circ$ denotes the Stratonovich product, and represents physically the mean velocity of the process at the point $x$. The Stratonovich product is used instead of the It\^o product to ensure that, as $T\ra\infty$, $J_T$ converges in probability to the stationary current $J_{F,\pi}$ of the diffusion defined in \eqref{eqsde2} \cite{barato2015}. Therefore, in that limit, we have $J_T\ra J_{F,\pi}$ in addition to $\rho_T\ra \pi$ in probability, so that the most probable value of the couple $(\rho_T,J_{T})$ is $(\pi,J_{F,\pi})$. 

The likelihood of fluctuations around this concentration point is quantified, similarly to $A_T$ and $\rho_T$, by an LDP whose rate function is known to be
\be
\Itwofive_C(\rho,j) = \frac{1}{2}\int_{\reals^d} (j-J_{F,\rho})\cdot (D\rho)^{-1}(j-J_{F,\rho})\, dx
\label{eqldp25}
\ee
if $\rho$ is a normalised density and $\nabla\cdot j=0$ \footnote{The expression \eqref{eqldp25} is valid for any invertible noise matrix $D$, so not necessarily the scalar $D$ assumed here. Moreover, $\Itwofive_C(\rho,j)=\infty$ if $\nabla\cdot j\neq 0$, so that current fluctuations that are not sourceless have a probability that decays faster than exponentially with $T$ \cite{barato2015}.}. This means that the joint probability of observing a density fluctuation $\rho$ away from $\pi$ together with a current fluctuation $j$ away from $J_{F,\pi}$ decays exponentially as $T\ra \infty$ with a rate given by the function \eqref{eqldp25}, which is explicit in $\rho$ and $j$. The term $J_{F,\rho}$ in $\Itwofive_C$ is given similarly as in \eqref{eqcurr1} by
\be
J_{F,\rho} =F\rho -\frac{D}{2}\nabla\rho = (-\nabla U+C)\rho -\frac{D}{2}\nabla \rho,
\label{eqinstcurr1}
\ee 
and is interpreted as an instantaneous current that ``sustains'' the density fluctuation $\rho$. For $\rho\neq\pi$, this current is obviously not the stationary current with respect to $F$ and $D$, so that $\nabla\cdot J_{F,\rho}\neq 0$. Also note that, despite the quadratic form of $\Itwofive_C$, the joint fluctuations of $\rho_T$ and $J_T$ are not Gaussian, although the fluctuations of $j$ are Gaussian around $J_{F,\rho}$ conditionally on $\rho_T=\rho$.

From the rate function of $(\rho_T,J_T)$, we can express the rate function of $\rho_T$ alone using the contraction principle as
\be
\Itwo_C(\rho) = \inf_{j:\nabla\cdot j=0} \Itwofive_C(\rho,j).
\label{eqcont25to2}
\ee
This corresponds to marginalising $J_T$ in the joint LDP of $(\rho_T,J_T)$ to obtain the LDP of $\rho_T$ only. By further contracting on $\rho$, as in \eqref{eqcont1}, we can also express the rate function of $A_T$ as
\be
I_C(a) = \inf_{(\rho,j):\tA(\rho)=a,\nabla\cdot j=0} \Itwofive_C(\rho,j).
\label{eqcont25}
\ee
This is the level-2.5 representation of the level-1 rate function. For ergodic diffusions, it is known that the solution $(\rho^*,j^*)$ of this constrained minimization is unique and can be interpreted as the stationary density and current of a controlled diffusion, called the driven or effective process \cite{jack2010b,chetrite2013,chetrite2014,jack2015,chetrite2015}, associated with a given fluctuation $A_T=a$.

It is beyond the scope of this paper to explain this interpretation in detail; see \cite{chetrite2014}. For our purposes, we only need four simple but important results related to the driven process:

1- The solution $(\rho^*,j^*)$ of the constrained variational problem \eqref{eqcont25} represents the most probable density and current fluctuations of the process $X(t)$ conditionally on observing the fluctuation $A_T=a$ \cite[Sec.~3]{chetrite2015}. 

2- Following the start of the section, we can identify a unique diffusion that has $\rho^*$ and $j^*$ as its stationary density and stationary current. This diffusion is the driven process having $A_T=a$ as its typical estimator value \cite[Sec.~3]{chetrite2015}. For the original, nonreversible diffusion defined in \eqref{eqsde2}, the typical value $a^*$ arises from $\pi$ and $J_{F,\pi}\neq 0$, while for the reversible diffusion \eqref{eqsde1} it arises from the same $\pi$ but $J_{F,\pi}=0$. 

3- For estimators or observables $A_T$ that are purely additive in time, the driven process is always a gradient perturbation of the original diffusion considered, having the same noise as the original diffusion \cite[Sec.~5.5]{chetrite2014}. For the nonreversible diffusion defined in \eqref{eqsde2}, the driven process is thus a new diffusion $\tX(t)$ governed by the SDE
\be
d\tX(t) = [F(\tX(t)) +\nabla\phi(\tX(t)) ]dt+\sqrt{\frac{2}{\beta}}\, dW(t), 
\label{eqdriven1}
\ee
where $\phi$ is function determined from a spectral problem related to the generator of $X(t)$ \cite[Sec.~4]{chetrite2014}.

4- The previous point implies that fluctuations of $A_T$ for a gradient diffusion are always created by another (driven) gradient diffusion having zero stationary current. This yields the variational representation of $I_0(a)$ in \eqref{eqcont1}, with the explicit rate function in \eqref{eqldp2}, as it can be checked that
\be
\Itwo_0(\rho) =\Itwofive_0(\rho,0)
\ee 
when $j^*=0$ \cite[App.~A.2]{chetrite2015}.

%%%%%%%%%%%%%%%%%%%%%%%%%%%%%%%%%%%%%%%%%%%%%%%%%%%%%%%%%%%
\section{Results}
\label{secres}

We are now ready to prove the large deviation inequalities obtained by Rey-Bellet and Spiliopoulos using the results listed above and the level-2.5 representation \eqref{eqcont25to2} of the level-2 rate function. The first step is to decompose the instantaneous current $J_{F,\rho}$ associated with the drift $F$ and density fluctuation $\rho$ as
\be
J_{F,\rho} = J^\rev_{F,\rho}+  J^\irr_{F,\rho},
\label{eqcurrdec1}
\ee
where
\be
J_{F,\rho}^\rev = -\nabla U\rho -\frac{D}{2}\nabla \rho = \frac{D}{2}\rho\nabla \ln \frac{\pi}{\rho}
\label{eqcurrrev1}
\ee
and
\be
J_{F,\rho}^\irr =J_{F,\rho}-J_{F,\rho}^\rev= C\rho. 
\label{eqcurrirr1}
\ee
This decomposition is natural in the context of nonequilibrium systems \cite{risken1996} and is justified here by noting that $-\nabla U$ is the reversible part of the drift, while $C$ is the irreversible part, giving rise in the stationary state to the non-zero current $J_{F,\pi} =  J_{F,\pi}^\irr = C\pi$. Note again that $J_{F,\rho}$ is not a stationary nor a sourceless current, in general, so neither $J_{F,\rho}^\rev$ nor $J_{F,\rho}^\irr$ has zero divergence, except for $\rho=\pi$. This is important to keep in mind.

Substituting \eqref{eqcurrdec1} in \eqref{eqldp25}, we now obtain
\be
\begin{split}
\Itwofive_C(\rho,j) &= \frac{1}{2}\snorm{ J_{F,\rho}^\rev}^2 +\frac{1}{2}\snorm{ j-J_{F,\rho}^\irr}^2\\
 &\qquad -\sprod{ (j-J_{F,\rho}^\irr)}{ J_{F,\rho}^\rev}
\end{split}
\label{eqmain1}
\ee
using the weighted scalar product
\be
\sprod{a}{b}=  \int_{\reals^d} a(x)\cdot (\rho(x) D)^{-1}b(x) \, dx
\ee
and $\snorm{a}^2=\sprod{a}{a}$ as the corresponding weighted norm. It can be checked with \eqref{eqcurrrev1} that the first term on the right-hand side of \eqref{eqmain1} is the level-2 rate function $\Itwo_0(\rho)$, shown in \eqref{eqldp22}, so we have in fact
\be
\begin{split}
\Itwofive_C(\rho,j) &= \Itwo_0(\rho)+\frac{1}{2}\snorm{j-J_{F,\rho}^\irr}^2\\
 &\qquad -\sprod{(j-J_{F,\rho}^\irr)}{J_{F,\rho}^\rev}.
\end{split}
\label{eqmain2}
\ee
This also follows since $\Itwo_0(\rho) = \Itwofive_0(\rho,0)$, as noted in the previous section, and $J_{F,\rho}^\irr =0$ for $C=0$. Evaluating the above expression at the minimizer $j^*$ of the contraction principle \eqref{eqcont25to2}, which connects the level 2.5 to the level 2, we then obtain
\be
\begin{split}
\Itwo_C(\rho) &= \Itwo_0(\rho)+\frac{1}{2}\snorm{j^*-J_{F,\rho}^\irr}^2\\
 &\qquad -\sprod{(j^*-J_{F,\rho}^\irr)}{J_{F,\rho}^\rev}.
\end{split}
\label{eqmain3}
\ee

At this point, we use \eqref{eqcurrrev1} to write
\be
\sprod{j^*}{J_{F,\rho}^\rev} = \frac{1}{2}\int_{\reals^d}dx\, j^*\cdot \nabla\ln\frac{\pi}{\rho},
\ee
which vanishes for all $\rho$ by the divergence theorem and the fact that $\Itwofive_C(\rho,j)$ is defined for sourceless current fluctuations satisfying $\nabla \cdot j^*=0$. As a result, $j^*$ is always orthogonal to $J_{F,\rho}^\rev$ with respect to the weighted scalar product:
\be
\sprod{j^*}{J_{F,\rho}^\rev}=0.
\label{eqortho1}
\ee 
The same orthogonality applies to $J_{F,\rho}^\irr$ and $J_{F,\rho}^\rev$, as can be checked from the definition of these currents and the fact that
\be
J_{F,\rho}^\irr=\frac{\rho}{\pi} J_{F,\pi}^\irr,
\ee 
so that, by the divergence theorem and the divergencelessness of $J_{F,\pi}^\irr$, we obtain
\be
\sprod{J_{F,\rho}^\irr}{J_{F,\rho}^\rev}=0
\label{eqortho2}
\ee
for all $\rho$. Thus, the second term in \eqref{eqmain3} vanishes, leaving
\be
\Itwo_C(\rho) = \Itwo_0(\rho)+\frac{1}{2}\snorm{j^*-J_{F,\rho}^\irr}^2.
\label{eqmain4}
\ee
Since the second term on the right-hand side is positive, we finally recover the inequality \eqref{eqldbound2}. The same reasoning applied to \eqref{eqcont25} yields the inequality \eqref{eqldbound1} for the rate functions of $A_T$ at the level 1 of large deviations.

We contend that this derivation is simpler than the one found in \cite{bellet2015}, although it requires, arguably, more background material on level-2.5 large deviations. An advantage of the level 2.5 approach is that it provides a physical interpretation of the large deviation inequalities: fluctuations of $\rho_T$ are less likely in the nonreversible system compared to the reversible one because these fluctuations are accompanied in general by current fluctuations that incur an additional large deviation cost in the rate function. By comparison, density fluctuations always arise in the reversible system with $j^*=0$, as mentioned before, since reversible processes have no current for a fixed density, stationary or fluctuating. The same applies to the fluctuations of $A_T$, since this random variable is a contraction of $\rho_T$, so its large deviations are determined from those of  $\rho_T$.

These results and interpretations apply beyond the estimation problem to any ergodic diffusion processes, which means that they can be used to understand how density and current fluctuations arise in more general processes, be they used as models of nonequilibrium systems or for simulations. The crucial point in our analysis is to be able to express the drift $F$ similarly to the current decomposition \eqref{eqcurrdec1} as
\be
F= F_\rev+F_\irr,
\ee 
where
\be
F_\rev = \frac{D}{2}\nabla\ln \pi
\ee
is the reversible part of the drift associated with the stationary distribution $\pi$, which is not necessarily a Gibbs distribution. The current associated with the reversible part of the drift is such that $J_{F_\rev,\pi}=0$, while $F_\irr = F-F_\rev$ is the remaining, irreversible part of the drift such that $J_{F_\irr,\pi} = J_{F,\pi}\neq 0$. Following the reasoning above, we can then write
\be
\Itwofive_\irr(\rho,j) = \Itwofive_\rev(\rho,0)+\frac{1}{2}\snorm{j-J_{F,\rho}^\irr}^2,
\ee
where $\Itwofive_\irr$ is the level-2.5 rate function of the full, nonreversible diffusion with drift $F$, while $\Itwofive_\rev$ is the corresponding rate function of the reversible diffusion obtained with $F_\rev$ only. Thus, we see that the large deviation cost of producing a joint density and current fluctuation in a nonreversible diffusion is the cost of producing the density fluctuation in a reversible diffusion, which has the same stationary distribution but no current, plus a quadratic cost involving the irreversible current $J_{F,\rho}^\irr=F_\irr \rho$.

The orthogonality conditions that we have derived for the particular SDE \eqref{eqsde2} can also be applied in a more general way to any ergodic diffusion with drift $F$, provided that we define $J_{F,\rho}^\rev$ with the second equality in \eqref{eqcurrrev1}, that is,
\be
J_{F,\rho}^\rev = \frac{D}{2}\rho\nabla \ln \frac{\pi}{\rho}=  F_\rev\rho-\frac{D}{2}\nabla \rho = J_{F_\rev,\rho}.
\ee 
Thus, the first orthogonality condition \eqref{eqortho1} holds in general for this current, while the second condition in \eqref{eqortho2} holds for
\be
J_{F,\rho}^\irr=J_{F,\rho}-J_{F,\rho}^\rev=F_\irr\rho.
\ee
Moreover, note in the case of \eqref{eqortho1} that $j^*$ can be replaced by any sourceless current, so this orthogonality condition can be generalized to
\be
\sprod{j}{J_{F,\rho}^\rev}=0
\label{eqortho3}
\ee
for any $\rho$ and $j$ such that $\nabla\cdot j=0$. This result is potentially useful for simplifying analytical or numerical calculations at the level 2.5 of large deviations, since it constrains the class of current fields that are possible solutions to either the contraction \eqref{eqcont25to2} or \eqref{eqcont25}. 

To close this section, we return to the original SDEs \eqref{eqsde1} and \eqref{eqsde2} to discuss a few technical but important results. First, note from \eqref{eqmain4} that
\be
\Itwo_C(\rho) = \Itwo_0(\rho)
\label{eqeq1}
\ee 
if and only if $j^* = J_{F,\rho}^\irr=C\rho$, which leads with $\nabla\cdot j^*=0$ to $\nabla\cdot (C\rho)=0$. We know from \eqref{eqerg1} that the latter equation admits the solution $\rho=\pi$, so that
\be
\Itwo_C(\pi) = \Itwo_0(\pi)=0
\ee 
and, by contraction,
\be
I_C(a^*) = I_0(a^*) = 0.
\ee
Extra conditions on $C$ and the observable $A_T$ are required to show that having $C\neq 0$ leads to a strict inequality, $I_C(a)>I_0(a)$, away from $a^*$; see \cite[Thm.~2.4]{bellet2015}.

The main result in \eqref{eqmain4} can also be used to understand the result shown in \eqref{eqrbs1}. Recall from the previous section that fixing the value of $(\rho_T,J_T)$ identifies the driven process in a unique way and that this process is known to be a gradient perturbation of the original nonreversible diffusion, as shown in \eqref{eqdriven1}. As a result, we can write
\be
\label{eqcurropt}
\begin{split}
j^*  &= (-\nabla U+C+\nabla\phi)\rho-\frac{D}{2}\nabla\rho\\
& = \rho(-\nabla U+C+\nabla \phi+\nabla V),
\end{split}
\ee
where $V$ is again the potential associated with the density fluctuation $\rho$. Inserting this current in \eqref{eqmain4} with $J_{F,\rho}^\irr=C\rho$ then leads to \eqref{eqrbs1} with $\psi = \phi+V$, where $V$ is as before the potential associated with $\rho$ defined in \eqref{eqVpot1}. Thus, the quadratic cost in \eqref{eqrbs1} involving $\psi$ is nothing but the current cost. Moreover, the abstract equation shown in \eqref{eqrbscons1} is nothing but the current constraint $\nabla\cdot j^*=0$ ensuring that $j^*$ is a stationary current for the driven process.

We refer again to \cite{chetrite2013,chetrite2014,jack2015,chetrite2015} for more information about the properties and interpretation of the driven process. Incidentally, the fact that this process is a gradient perturbation of the process $X(t)$ with drift $F=-\nabla U+C$ can be inferred from the contraction formula \eqref{eqcont25to2} or \eqref{eqcont25}, as explained in the Appendix A.1 of \cite{chetrite2015}. This property of the driven process only holds for additive functionals, such as $\rho_T$ or $A_T$; for more general functionals involving $X(t)$ and the increments of $X(t)$, the driven process is in general a nonreversible process with both gradient and non-gradient terms added to $F$ \cite{chetrite2013,chetrite2014,jack2015,chetrite2015}.

Finally, note that if we fix $j=0$ in the contraction \eqref{eqcont25to2}, relating the level 2.5 to the level 2, then we obtain the inequality
\be
\Itwo_C(\rho) =\Itwofive_C(\rho,j^*) \leq \Itwofive_C(\rho,0),
\ee
since $j^*\neq 0$ in general, so that
\be
\Itwo_C(\rho)\leq\frac{1}{2}\snorm{J_{F,\rho}}^2.
\label{eqcurrbound1}
\ee
Therefore, a density fluctuation is less likely to be produced in a nonreversible diffusion by forcing the current to be zero, as for a reversible diffusion, than by allowing $j$ to attain its optimal value $j^*$, corresponding again as the most probable current of $X(t)$ conditionally on observing $\rho_T=\rho$. This is clear if we consider ``small'' fluctuations of $\rho_T$ around the stationary distribution $\pi$. Then we expect that $j^*$ should differ only slightly from the stationary current $J_{F,\pi}$, so having $J_T=0$ is very unlikely. Combining this result with the inequality \eqref{eqldbound2}, which can be re-expressed as
\be
\Itwo_C(\rho) \geq  \frac{1}{2}\snorm{J_{F,\rho}^\rev}^2,
\ee
then shows that a density fluctuation is \emph{more} likely to appear in a reversible system than in a nonreversible system (lower bound), although it is \emph{less} likely to appear in a nonreversible system that behaves at a fluctuation level like a reversible system (upper bound). Speaking in terms of large deviation cost, this means that it costs less to produce a density fluctuation in a reversible system than in a nonreversible system, although it costs more to produce that density fluctuation in a nonreversible system forced to have $J_T=0$.

Note, at a more mathematical level, that if the divergence operator ($\nabla\cdot$) is assumed to be invertible, then the upper bound in \eqref{eqcurrbound1} can formally be written as
\be
\Itwo_C(\rho)\leq\frac{1}{2}\|\nabla\cdot J_{F,\rho}\|_{-\nabla\rho D\nabla}^2.
\label{eqmftbound1}
\ee
Moreover, if we assume the same invertibility condition and use the sourceless condition $\nabla\cdot j=0$ in \eqref{eqldp25}, then we find that $\Itwo_C(\rho)$ is given by the same expression on  the right-hand side above, but now as an equality. This shows that the inequality \eqref{eqcurrbound1} results in general from the non-invertibility of $\nabla\cdot$.

%%%%%%%%%%%%%%%%%%%%%%%%%%%%%%%%%%%%%%%%%%%%%%%%%%%%%%%%%%%
\section{Applications}

We illustrate in this section our results using a version of the Ornstein--Uhlenbeck process in two dimensions and the simple diffusion on the circle, showing how nonreversible drift terms accelerate the convergence of estimators. The results are presented for different observables both at the level 1 of large deviations and the level 2 when the latter is explicit.

\subsection{Ornstein--Uhlenbeck process}

The first model that we consider is defined by
\be
dR(t) = -R(t) dt + d W(t) ,
\label{eq:OU2DRev}
\ee
where $R(t) = (X(t),Y(t))^\transp \in\reals^2$ is the state vector of the process, with $\transp$ denoting the transpose, and $W(t) = (W_x(t),W_y(t))^\transp$ is a Brownian motion in $\reals^2$ with independent components. This version of the Ornstein--Uhlenbeck process is reversible, as its drift can be derived from the potential
\be
U(x,y) = \frac{x^2 + y^2}{2}.
\ee
To make it nonreversible, we add the curl drift $C(x,y) = (y,-x)^\transp$ so as to modify \eqref{eq:OU2DRev} to
\be
\begin{pmatrix} dX(t)\\ dY(t) \end{pmatrix} = 
\begin{pmatrix} -X(t) + Y(t)  \\ -X(t) -Y(t) \end{pmatrix} dt 
+ \begin{pmatrix} d W_x(t)\\ 	dW_y(t) \end{pmatrix},
\label{eq:OU2DNonRev}
\ee
that is,
\be
dR(t) = -M R(t) dt +dW(t)
\ee
in vector form, where
\be
M = 
\left(
\begin{array}{cc}
1 & -1 \\
1 & 1
\end{array}
\right).
\ee
Since $C$ satisfies the condition \eqref{eqerg1}, the invariant density for both processes has the Gibbs (Gaussian) form \eqref{eqgibbs1},
\be
\pi(x,y) = \frac{e^{-2U(x,y)}}{N},
\label{eq:OU2DInv}
\ee
where $N$ is a normalization constant, while the stationary current of the nonreversible process is given by
\be
J_{F,\pi} (x,y) = C(x,y) \pi(x,y),
\label{eq:OU2DStatCurr}
\ee
in agreement with \eqref{eqcurr1}.

Various observables can be chosen to illustrate the accelerated convergence of the nonreversible SDE. Here, we simply consider
\be
A_T = \frac{1}{T} \int_0^T X(t) dt 
\label{eq:OU2DObs}
\ee
and proceed to find the level-1 rate function of this real random variable for both the reversible and nonreversible SDEs. This can be done, in principle, using the contraction formula \eqref{eqcont25}, which expresses $I(a)$ as a contraction of the level-2.5 rate function. For this example, however, it is easier to obtain $I(a)$ directly at the level 1 of large deviations by calculating the scaled cumulant generating function (SCGF) $\lambda(k)$ of $A_T$ as the dominant eigenvalue of the spectral problem
\be
\label{eq:OU2DSpectral}
\mathcal{L}_k r_k(x,y) = \lambda(k) r_k(x,y) , 
\ee
where $k$ is a real parameter, $\mathcal{L}_k$ is a linear operator called the tilted generator, and $r_k(x,y)$ is the eigenfunction, defined on $\reals^2$, associated with $\lambda(k)$. Solving this spectral problem, we then obtain the rate function $I(a)$ of $A_T$ using the Legendre transform
\be
I(a) = k(a) a-\lambda(k(a)),
\ee
where $k(a)$ is fixed by 
\be
a = \lambda'(k(a)).
\label{eq:OU2DDuality}
\ee
We refer to \cite{touchette2017} for a presentation of these results and the conditions underlying them. For the reversible process \eqref{eq:OU2DRev}, the tilted generator reads
\be
\label{eq:OU2DTiltRev}
\mathcal{L}_k = -\nabla U \cdot \nabla + \frac{\Delta}{2} + k x ,
\ee
where $\nabla = (\p/\p x, \p/\p y)^T$ and $\Delta = \p^2/\p x^2 + \p^2 /\p y^2$ is the Laplacian in $\reals^2$, whereas for the nonreversible process \eqref{eq:OU2DNonRev}, we have
\be
\label{eq:OU2DTiltnonRev}
\mathcal{L}_k = (-\nabla U + C) \cdot \nabla + \frac{\Delta}{2} + k x .
\ee

The spectral problem associated with the tilted generator of the reversible process can easily be solved, giving
\be
\lambda_0(k) = \frac{k^2}{2}
\label{eq:OU2DSCGFRev}
\ee
with $r_{0,k}(x,y) = e^{kx}$, where the subscript $0$ indicates, as before, that we are dealing with the reversible process. From the Legendre transform \eqref{eq:OU2DSCGFRev}, we then obtain
\be
I_{0}(a) = \frac{a^2}{2},
\label{eq:OU2DRateRev}
\ee
showing that the large deviations of $A_T$ are Gaussian around $E[A_T]=a^*=0$, with asymptotic variance $\sigma_{0}^2=1$ given by \eqref{eqasympvarld1}. For the nonreversible process, the spectral problem can also be solved exactly and yields
\be
\lambda_{C}(k) = \frac{k^2}{4}
\label{eq:OU2DSCGFNonRev}
\ee
with
\be
r_{C,k}(x,y) = e^{\frac{k}{2}(x+y)}.
\label{eq:OU22RightNonRev}
\ee
In this case, the rate function obtained from the Legendre transform of $\lambda_C(k)$ is
\be
I_{C}(a) = a^2,
\label{eq:OU2DRateNonRev}
\ee
so that $\sigma_{C}^2=1/2$, confirming the inequalities \eqref{eqldbound1} and \eqref{eqvarbound1} for the rate functions and asymptotic variances, respectively. Hence, the convergence of $A_T$ to $a^*=0$ is faster when simulating the nonreversible process than when simulating the reversible one.

\begin{figure*}[t]    
\includegraphics[width=\linewidth]{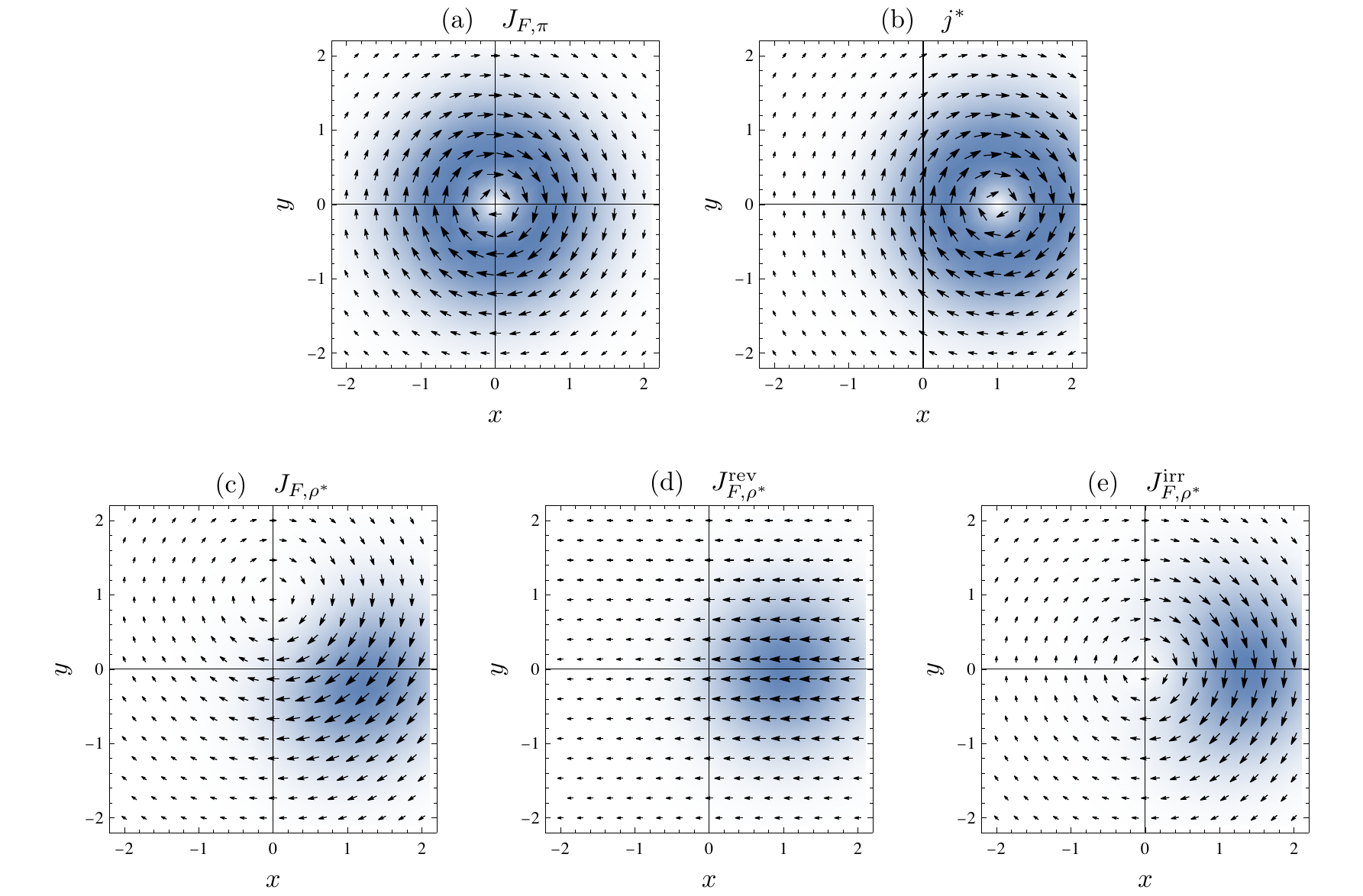}
\caption{(Color online) Vector plots of stationary and fluctuating currents associated with the Ornstein--Uhlenbeck model. The density plot shows the magnitude of the corresponding current field. (a)  Stationary current $J_{F,\pi}$. (b) Stationary current $j^*=J_{F_{C,k(a)},\rho^*}$ of the driven process corresponding to the fluctuation $A_T=a$. (c) Instantaneous current $J_{F,\rho^*}$. (d) Reversible part of the instantaneous current. (e) Irreversible part of the instantaneous current. Parameters: $k=2$ corresponding to $a=1$.}
\label{fig:OUChart}
\end{figure*}

To understand physically why there is faster convergence, we now consider the driven process describing how the fluctuations of $A_T$ are created by density and current fluctuations. In the reversible case, we find from \eqref{eqdriven1} and $\phi=\frac{2}{\beta}\ln r_k$ that the driven process is a linear (affine) process described by the modified drift
\be
F_{0,k}(x,y) = -\begin{pmatrix} x\\ y \end{pmatrix} + \begin{pmatrix} k\\ 0 \end{pmatrix},
\ee
whereas the modified drift associated with the nonreversible process is
\be
F_{C,k}(x,y) = -M\begin{pmatrix} x\\ y \end{pmatrix} +\frac{1}{2} \begin{pmatrix} k\\ k \end{pmatrix}.
\ee
These drifts are parameterized by $k$; to relate them to a given fluctuation $A_T=a$, we need to use the duality relation \eqref{eq:OU2DDuality} of the Legendre transform to obtain
\be
F_{0,k(a)}(x,y) = \begin{pmatrix} -x+a\\ -y \end{pmatrix}
\label{eqdrivendriftrev1}
\ee
and
\be
F_{C,k(a)}(x,y) = \begin{pmatrix} -x+y+a\\ -x-y+a \end{pmatrix}.
\label{eqdrivendriftnonrev1}
\ee
From this, we see that a fluctuation $A_T=a$ is created in the reversible process by translating the attractor of that process on the $x$-axis to $a$, effectively moving the stationary density $\pi$ to a new, fluctuating density $\rho^*$ centered at $(a,0)$, that is,
\be
\rho^*(x,y) = \pi(x-a,y).
\label{eqdrivendensity1}
\ee
In this way, the process spends most of its time around $(a,0)$, leading to $A_T\ra a=E_{\rho^*}[X]$ in the ergodic limit. In this case, we also have $j^*=0$ for all $a$, since the driven process associated with the reversible process is known to be reversible. Thus, fluctuations of $A_T$, which involves only the component $X(t)$ of the process, are created by translating the stationary density on the $x$-axis, while keeping the zero current.

For the nonreversible process, the stationary density $\pi$ is also modified to $\rho^*$, shown in \eqref{eqdrivendensity1}, since the drift \eqref{eqdrivendriftnonrev1} is still only a translation in the $x$-direction of the original drift in \eqref{eq:OU2DNonRev}, but the stationary current is now modified as a result of this density fluctuation to
\be
j^*=J_{F_{C,k},\rho^*}=\left[ - \nabla U + C + \frac{1}{2} \begin{pmatrix} k  \\ k \end{pmatrix} \right] \rho^* - \frac{1}{2} \nabla \rho^*. 
\label{eqdrivencurr1}
\ee
It can be checked, by changing the parameterization to $a$, that this is equivalent to 
\be
j^*(x,y)=J_{F,\pi}(x-a,y) = \rho^*(x,y) \begin{pmatrix} y\\ -x+a \end{pmatrix}
\label{eqdrivencurr2s}
\ee
so the current underlying the fluctuation $A_T=a$ is an $x$-translation of the stationary current, similarly to the density, as shown in Fig.~\ref{fig:OUChart}(a)-(b) for $a=1$. This modified current is responsible for the added large deviation cost, which makes the fluctuation less likely in the nonreversible process compared with the reversible process.

From the known density $\rho^*$, we can also find explicit expressions for the instantaneous current $J_{F,\rho^*}$ associated with $A_T=a$, as well as for its reversible and irreversible components, defined by \eqref{eqcurrrev1} and \eqref{eqcurrirr1}, respectively. The results are
\be
\begin{split}
J_{F,\rho^*}(x,y) &= \rho^*(x,y) \begin{pmatrix} y-a\\ -x \end{pmatrix}\\
J_{F,\rho^*}^\rev(x,y) &= \rho^*(x,y) \begin{pmatrix} -a\\ 0 \end{pmatrix}\\
J_{F,\rho^*}^\irr(x,y) &= \rho^*(x,y) \begin{pmatrix} y\\ -x \end{pmatrix}.
\end{split}
\ee
These currents are illustrated in Fig.~\ref{fig:OUChart}(c)-(e) for $a=1$. Following Gauss's flux theorem, it is clear from the plots shown, especially from that of $J_{F,\rho^*}^\rev$, that the instantaneous currents are not sourceless in general. The orthogonality of $j^*$ and $J_{F,\rho^*}^\rev$ is not as obvious visually, as it is defined with respect to the weighted scalar product, but can be checked explicitly from the expressions above. The same applies for the orthogonality of $J_{F,\rho^*}^\rev$ and $J_{F,\rho^*}^\irr$.

These results can be generalized slightly by replacing $C$ with $\alpha C$, using $\alpha\in\reals$ as a parameter to continuously go from the reversible case ($\alpha=0$) to the nonreversible case ($\alpha\neq 0$). It can be checked that the spectral calculation with $\alpha$ yields 
\be
\lambda_{\alpha C}(k) = \frac{k^2}{2(1+\alpha^2)}
\ee
and
\be
I_{\alpha C}(a) = \frac{(1+\alpha^2) a^2}{2},
\ee
leading to 
\be
\sigma^2_{\alpha C} = \frac{1}{1+\alpha^2}
\ee
for the asymptotic variance. This recovers our previous results for $\alpha=1$ and shows that there is acceleration as soon as the curl field is switched on with $\alpha\neq 0$.

\begin{figure}[t]
\includegraphics{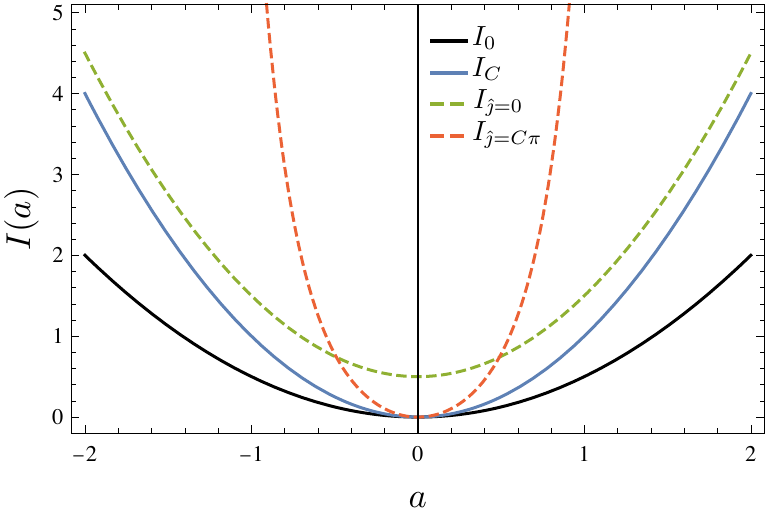}
\caption{(Color online) Comparison of the rate functions $I_0(a)$ and $I_C(a)$ obtained for the reversible and nonreversible Ornstein--Uhlenbeck processes, respectively. $I_{\hj=0}$ and $I_{\hj = C\pi}$ are upper bounds on $I_C(a)$ obtained from the sub-optimal currents $\hj=0$ and $\hj=C\pi$, respectively.}
\label{fig:bounds}
\end{figure}

Note that all of these results are explicit because the driven process retains the linear form of the Ornstein--Uhlenbeck process when considering a linear observable. For more general processes and observables, the spectral problem \eqref{eq:OU2DSpectral} is unlikely to be solvable exactly, in which case we may try to approximate the result of the contraction in \eqref{eqcont25to2} using sub-optimal currents $\hj\neq j^*$, which yield upper bounds for the rate function. Three choices of currents are worth mentioning. First, we can use $\hj = 0$ to obtain the upper bound
\be
I_{C}(a) \leq \Itwofive_C(\rho^*,0) = a^2+\frac{1}{2}
\label{eqbound1}
\ee 
which corresponds to the upper bound \eqref{eqcurrbound1} with $\rho^*$. Second, we can choose $\hj = C\rho^*$, which, as we know, is not the true optimal fluctuating current $j^*$. However, since this current is not sourceless, it is not valid for the contraction. Instead, we can use $\hj = C\pi$ as an obvious third choice, which is divergenceless, to obtain the upper bound
\be
I_C(a) \leq  \Itwofive_C(\rho^*,C\pi) = \frac{a^2}{2}+e^{a^2}(a^2\cosh a^2+\sinh a^2).
\label{eqbound2}	
\ee
The two bounds \eqref{eqbound1} and \eqref{eqbound2} are compared in Fig.~\ref{fig:bounds} with the rate function $I_C(a)$ of the nonreversible process, as well as with the rate function $I_0(a)$ of the reversible process, lying below $I_C(a)$.

\subsection{Diffusion on the circle}

We revisit as our second example the Brownian motion on the circle, studied by Rey-Bellet and Spiliopoulos \cite{bellet2015} in the context of the acceleration problem at the level 2 of large deviations. We briefly show for this model how the level-2 rate function is obtained from the level-2.5 rate function, and consider a new type of acceleration obtained by changing the noise amplitude. 

The simple Brownian motion on the circle is defined by the SDE
\be
d \theta(t) = \sqrt{D}\, dW(t) \!\! \mod 2\pi,
\label{eq:CircleSDErev}
\ee
where $W(t)$ is the usual Brownian motion on $\reals$, so that $\theta(t) \in [0,2\pi)$, and $D$ is a real, positive constant determining the noise intensity. This motion is forced in the simplest way by adding a constant drift $C\in\reals$, yielding
\be
d \theta(t) = C dt + \sqrt{D}\, dW(t) \!\! \mod 2\pi,
\label{eq:CircleSDEnonrev}
\ee
as a nonreversible version of \eqref{eq:CircleSDErev}, which preserves the constant stationary density $\pi(\theta) = 1/(2\pi)$ \footnote{Note that we use $\pi$ at this point to mean the stationary density and the constant $\pi$.}. The stationary current $J_{C,\pi}(\theta) $ of the drifted motion is also constant and equal to $C/(2\pi)$. 

The level-2.5 rate function simplifies for the drifted Brownian motion because of the one-dimensional sourceless condition which implies that currents have to be constant. Hence, \eqref{eqldp25} takes the form
\be
\label{eq:CircleLD25}
\Itwofive_C(\rho,j) = \frac{1}{2} \snorm{J_{C,\rho}}^2 + \frac{j^2}{2} \snorm{1}^2 - j\sprod{1}{J_{C,\rho}} ,
\ee
where $j\in\reals$ and the scalar product is performed on the circle. The minimization with respect to $j$ in \eqref{eqcont25to2} is straightforward and leads to the level-2 large deviation rate function
\be
\Itwo_C(\rho) = \frac{D}{8} \sprodd{\rho^{-1}}{\left( \frac{d \rho}{d \theta} \right)^2} + \frac{C^2}{2D} \left( 1 - \frac{4 \pi^2}{\sprodd{\rho^{-1}}{1}} \right)
\label{eq:CircleLD2}
\ee
with the optimal current
\be
j^*=\frac{2\pi C}{\langle \rho^{-1},1\rangle}.
\label{eq:currroptring}
\ee
Note that we use now the unweighted scalar product $\langle\cdot,\cdot\rangle = \langle\cdot,\cdot\rangle_{1}$ obtained with $\rho D=1$ to make the dependence on $D$ more explicit. 

The result in \eqref{eq:CircleLD2} recovers for $D=1$ the level-2 rate function found in Example 2.9 of \cite{bellet2015}: the first term on the right-hand side is the explicit rate function in the reversible case, as given by \eqref{eqldp2}, whereas the second term is the added contribution coming from the non-gradient forcing $C$. Acceleration follows from the latter term, since $\sprodd{\rho^{-1}}{1} \geq 4 \pi^2$ by the Cauchy-Schwarz inequality, so the inequality \eqref{eqldbound2} at the level 2 of large deviations holds, implying the inequality \eqref{eqldbound1} at the level 1 for any additive observables. The same inequality implies with \eqref{eq:currroptring} that $j^*\leq J_{C,\pi}$, which means that density fluctuations reduce the stationary current.

This calculation shows that statistical acceleration follows by changing the average speed of the Brownian motion. Since $\pi$ for this model is also independent of $D$, it is interesting to see whether there is acceleration by changing the noise intensity. To this end, we can take the first derivative of the level-2 rate function \eqref{eq:CircleLD2} with respect to $D$ for a given $\rho$ to obtain
\be
\frac{\p \Itwo_C(\rho)}{\p D} = \frac{1}{8}\left\langle \rho^{-1},\left(\frac{d\rho}{d\theta}\right)^2\right\rangle - \frac{C^2}{2D^2}\left(1-\frac{4\pi^2}{\langle\rho^{-1},1\rangle}\right).
\label{eq:CircleL2DerD}
\ee
This result is non-negative when
\be
D \geq 2 |C|\sqrt{\frac{1-\frac{4\pi^2}{\langle\rho^{-1},1\rangle}}{\left\langle\rho^{-1},\big(\frac{d\rho}{d\theta}\big)^2\right\rangle}}.
\label{eq:CircleNonNegativeLdDerD}
\ee
Consequently, if we choose $D$ large enough according to the inequality above, then $\Itwo_C(\rho)$ increases, implying that fluctuations around $\rho$ are exponentially suppressed. Hence, adding noise suppresses density fluctuations.

This is a counter-intuitive effect, which holds, interestingly, for any $D>0$ when there is no drift ($C=0$). This means that increasing the noise intensity in the simple Brownian motion always leads to a statistical acceleration of $\rho_T$ towards $\pi$ and, by contraction, a statistical acceleration of additive observables towards their ergodic values. The same holds when $C\neq0$ if we consider density fluctuations close to $\pi$. In this case, the expansion of \eqref{eq:CircleL2DerD} around $\pi$ using the perturbation $\rho=\pi+\delta \rho$ with
\be
\int_0^{2\pi} \delta \rho(\theta) d\theta = 0,
\label{eq:CirclePerturb}
\ee
gives
\be
\at{\frac{\p \Itwo_C(\rho)}{\p D}}{\rho=\pi+\delta\rho} = \frac{1}{8} \sprodd{\pi^{-1}}{ \left( \frac{d \delta \rho}{d \theta} \right)^2 } + O(\delta \rho^3).
\label{eq:CircleL2DerDExpans}
\ee
The right-hand side of above is obviously non-negative, so the level-2 rate function increases or stays the same around $\pi$ if we increase $D$, which means again that we can accelerate the statistical convergence of $\rho_T$ to $\pi$ by increasing the noise intensity.

This result, we should emphasize, is specific to additive observables having the form defined in \eqref{eqobs1}. If we consider more general observables contracted from $\rho_T$ and $J_T$ rather than just $\rho_T$, then the rate function is generally less steep if we increase the noise intensity. A case in point is the empirical velocity of the drifted Brownian motion, defined by
\be
B_T = \frac{1}{T}\int_0^T d\theta(t) = \int_0^{2\pi} J_T(\theta)\, d\theta,
\ee
which has the trivial rate function \cite{tsobgni2016}
\be
I_C(b) = \frac{(b-C)^2}{2D},
\ee
describing Gaussian fluctuations around the expected velocity $b^*=2\pi J_{C,\pi}=C$. Therefore, increasing the noise in this case increases rather than decreases the asymptotic variance. It is also clear that increasing the drift does not decrease the asymptotic variance.

%%%%%%%%%%%%%%%%%%%%%%%%%%%%%%%%%%%%%%%%%%%%%%%%%%%%%%%%%%%
\section{Conclusion}

We have used the theory of large deviations to explain why statistical estimators defined in the context of diffusion processes (viz., Langevin samplers) converge faster in time when adding irreversible component to the drift. The accelerated convergence comes, as we have shown, because the fluctuations of these estimators are exponentially suppressed in nonreversible processes, compared to reversible ones, due to current fluctuations being created in the former. We have studied these currents using the level 2.5 of large deviations and have shown that there is a most probable current fluctuation arising in conjunction with a given observable fluctuation. This current fluctuation can be interpreted as an optimal current field that yields the large deviation function of the observable of interest, as well as the stationary current of a modified diffusion, called the driven or effective process, which describes how fluctuations arise in general by means of modified densities and currents. 

These results provide a new interpretation of the work of Rey-Bellet and Spiliopoulos \cite{bellet2015,bellet2015b,bellet2016}, who established the accelerated convergence of nonreversible samplers by studying density fluctuations at the level 2 of large deviations. Our approach consists in adding current fluctuations in the analysis using the level 2.5 of large deviations in order to get a more complete and physical understanding of nonreversible samplers compared to reversible ones. This provides an application of the level 2.5 in statistical estimation, complementing the more physical applications that have been discussed up to now \cite{gingrich2016,pietzonka2015,gingrich2017,nardini2018,monthus2019,monthus2019b,monthus2021}.

We should mention that another study \cite{kaiser2017} has looked at the role of fluctuating currents in the acceleration problem, but has done so in the context of a large deviation formalism known as the macroscopic fluctuation theory (MFT) \cite{bertini2006,hurtado2014,bertini2015b}, in which the orthogonal decomposition of forces and currents that we have used also appears naturally \cite{kaiser2018,renger2018,renger2021,patterson2021}. The level 2.5 of large deviations looks similar in form to that theory \footnote{The expression on the right-hand side of \eqref{eqmftbound1}, for instance, is similar to the density rate function found in the MFT when fields are considered stationary; see, e.g., \cite[Eq.~(4.54)]{bouchet2016}.}, but the two are in fact different as they deal with different random variables and scaling limits: the level 2.5 deals, as we have seen, with the fluctuations of the empirical density and current of a single process in the long-time or ergodic limit, whereas the MFT is concerned with the occupation and current of many-particle Markov dynamics as a function of time in the limit where the number of particles goes to infinity, similarly to the thermodynamic limit of equilibrium systems. The correct formalism to use for studying the convergence of statistical estimators is the level 2.5 because that convergence is in the ergodic limit.

Future studies could focus on generalising our results to jump processes and Markov chains, which have also been covered by Rey-Bellet and Spiliopoulos at the level 2 of large deviations \cite{bellet2016}. The case of Markov chains evolving in discrete time is especially interesting, since it underlies many algorithms used in Monte Carlo simulations, in particular, the Metropolis algorithm. An obvious complication is that these algorithms often involve high-dimensional (discrete) state spaces on which currents, defined as a multi-dimensional matrices, cannot be visualised easily. However, the very presence of fluctuating currents could be sufficient to prove acceleration, as done here, starting from the known expression of the joint density-current (or density-flow) rate function of Markov chains or Markov jump processes \cite{barato2015}.

Another interesting problem is to study more general observables defined as contractions of the empirical current or the empirical flow in the case of jump processes \cite{barato2015}. We have shown with the example of the Brownian motion on the circle that adding a nonreversible drift does not necessarily decrease the asymptotic variance of current-like observables, but there might exist a class of such observables, or mixed observables obtained by contraction of the empirical density and current (or flow), for which the variance does decrease. 

Finally, it should be clear that, although we have focused on the simple case where $D$ is a scalar, much of our results can be generalized to a larger class of nonreversible diffusions that include different noise temperatures, as well as correlated and multiplicative noise \cite{bellet2016}. This is suggested by the way that we have written the level-2.5 rate function in \eqref{eqldp25} with the term $(\rho D)^{-1}$, which applies to any ergodic diffusions with drift $F$ and noise matrix $D$ assumed to be invertible. From this more general rate function, one can extend our results by noting that currents are defined not with the drift itself but with a modified drift that takes the multiplicative nature of the noise into account \cite{barato2015}. 

%%%%%%%%%%%%%%%%%%%%%%%%%%%%%%%%%%%%%%%%%%%%%%%%%%%%%%%%%%%
\begin{acknowledgments}
F.C. is grateful to the Institute of Mathematics and its Applications (Small Grant scheme) and the University of Nice for its hospitality and support during a visit to finalize this project. The research of R.C.\ is supported by the French National Research Agency through the projects QTraj (ANR-20-CE40-0024-01), RETENU (ANR-20-CE40-0005-01), and ESQuisses (ANR-20-CE47-0014-01).
\end{acknowledgments}

%%%%%%%%%%%%%%%%%%%%%%%%%%%%%%%%%%%%%%%%%%%%%%%%%%%%%%%%%%%
\bibliography{masterbib}

\end{document}